\documentclass[prb,amssymb,twocolumn]{revtex4}
\usepackage{dcolumn}
\usepackage{amsmath}
\usepackage{bm}

\begin{document}
\title{Resonance Raman scattering in semiconductor quantum
dots: Adiabatic vs. time-dependent perturbation theory}
\author{E. Men\'{e}ndez-Proupin}
\email{eariel@ff.oc.uh.cu}
\author{Nana Cabo-Bisset}
\affiliation{IMRE-Facultad de F\'{\i}sica, Universidad de La Habana, Vedado\\
10400, La Habana, Cuba}
\date{\today }

\begin{abstract}
The adiabatic theory of resonance one-phonon Raman scattering in
semiconductor nanocrystals is revised and extended with perturbative
non-adiabatic corrections, given by the Albrecht's B term. This theory is
confronted with the time-dependent perturbation approach, pointing at their
differences and similarities. It is shown that both theories are equivalent
in the limit of weak electron-phonon coupling and non-degenerate or
uncoupled resonant states. Evaluations of the A and B terms for the confined
LO phonon in CdSe and CdS nanocrystals are reported. These evaluations show
that the B term can usually be neglected.
\end{abstract}


\maketitle

\section{Introduction}

In the interpretation of resonant Raman scattering in quantum dots, two
theories have been mostly used. The first one is based on the Albrecht's
theory\cite{albrecht} of light scattering from small molecules, where the
excited levels that contribute to the Raman polarizability are considered as
vibron states in the adiabatic approximation. This theory also considered
perturbative non-adiabatic corrections, but these later have been neglected
in quantum dot studies \cite%
{klein,alivisatos,shiang,jung,scamarcio,cucl,cubr,krauss2}. Moreover, the
exciton-phonon coupling factor has been identified with the Huang-Rhys
factor, using this as a fitting parameter for the relative intensities of
different orders in Raman spectra. However, the calculations of the
Huang-Rhys parameter for intrinsic exciton states in PbS\cite{krauss2} and
CdSe\cite{nomura,chenjlum} nanocrystals have given values that are too small
compared with those needed to explain the experimental results of
multiphonon Raman scattering. Extrinsic mechanisms such as donor-like
exciton \cite{marini}, surface hole traps \cite{bawendi} or extra charges %
\cite{nomura,pokatilov01} have been invoked to resolve the discrepancy

A different model of Raman scattering in quantum dots was conceived from a
solid state point of view. In this approach, the Raman cross sections are
calculated from third or higher order time-dependent perturbation theory
(TDPT) \cite{chamb,multifonon,fedorov} and the intermediate virtual states
are considered as tensor products of electronic states, lattice
vibrations, and photons. A non-perturbative calculation of multiphonon Raman
spectra have been recently presented \cite{pokatilov01}. Not having
adjustable parameters this model has scarcely been used by experimentalists
to interpret their data. Moreover, up to now it is unclear the relation
between the TDPT and the Albrecht's theory.

The purpose of the present article is:\ (1) to establish the relation
between Albrecht's theory and TDPT, and (2) to investigate the importance of
non-adiabatic corrections within Albrecht's theory. The structure of the
paper is as fallows. First, we give an overview of both theoretical
approaches and show their interrelation. Next, we explain the calculation of
the Albrecht's A and B terms for semiconductor nanocrystals. Finally we
discuss the numerical results obtained for several types of nanocrystals and
present our conclusions. Several mathematical steps are given in the
appendixes.

\section{The theories}

\subsection{Time-dependent perturbation theory}

For a one-phonon Raman process the differential cross section is given by %
\cite{chamb} 
\begin{equation}
\frac{d^{2}\sigma }{d\Omega _{s}d\omega _{s}}=\frac{V^{2}\omega _{s}^{3}\eta
_{l}\eta _{s}^{3}}{4\pi ^{2}c^{4}\omega _{l}\hbar }\sum_{F}\left|
M_{FI}^{(1)}(p)\right| ^{2}\delta \left( \hbar \omega _{l}-\hbar \omega
_{s}-\hbar \omega _{p}\right) ,  \label{ec:seccion}
\end{equation}%
where $\eta _{l}$ ($\eta _{s}$) is the refraction index at the incident
(scattered) light frequency $\omega _{l}$ ($\omega _{s}$), $V$ is a
normalization volume of the radiation field, and $c$ is the velocity of
light in vacuum. The one-phonon transition amplitude $M_{FI}^{(1)}(p)$ can
be calculated by time dependent perturbation theory, considering the
unperturbed Hamiltonian as the sum of the Electronic, Lattice, and Radiation
operators 
\[
H_{0}=H_{E}+H_{L}+H_{R}, 
\]%
while the perturbation is the sum of Electron-Lattice, Electron-Radiation,
and Lattice-Radiation interactions (the last one is negligible in resonance
conditions) 
\[
H_{int}=H_{E-L}+H_{E-R}+H_{L-R}. 
\]%
The electron-lattice interaction can be expressed as 
\begin{eqnarray}
H_{E-L}&=&\sum_{\mu ,\mu ^{\prime },\nu }\left\langle \mu \left|
H_{E-L}^{+}(\nu )\right| \mu ^{\prime }\right\rangle \hat{D}_{\mu }^{\dagger
}\hat{D}_{\mu ^{\prime }}\hat{b}_{\nu }^{\dagger } \nonumber \\
&+&\sum_{\mu ,\mu ^{\prime
},\nu }\left\langle \mu \left| H_{E-L}^{-}(\nu )\right| \mu ^{\prime
}\right\rangle \hat{D}_{\mu }^{\dagger }\hat{D}_{\mu ^{\prime }}\hat{b}_{\nu
},
\end{eqnarray}
where $\hat{D}_{\mu ^{\prime }}$ ($\hat{D}_{\mu }^{\dagger }$) and $\hat{b}%
_{\nu }$\ ($\hat{b}_{\nu }^{\dagger }$) are annihilation (creation)
operators of electronic and vibrational excitations, respectively, while $%
H_{E-L}^{+}(\nu )$ and its Hermitian adjoint $H_{E-L}^{-}(\nu )$\ are
operators that act on the electronic system\cite{multifonon}. The operators $%
H_{E-R}$ and $H_{R-L}$ have a similar structure to that of $H_{E-L}$. The
phonon states\ created by $\hat{b}_{\nu }^{\dagger }$ are those of the
electronic ground state $G$. The electronic excitations $\mu \neq G$ can be
considered as electron-hole pairs and confined excitons. Hence, the
one-phonon Raman transition amplitude can be calculated by third order
perturbation theory as 
\begin{eqnarray}
M_{FI}^{(1)}(p)&=&\sum_{\mu _{1},\mu _{2}(\neq G)}\frac{\left\langle G\left|
H_{E-R}^{+}\right| \mu _{2}\right\rangle}
{\left( \hbar \omega _{l}-E_{\mu_{2}}-\hbar \omega _{p}+
i\Gamma _{\mu _{2}}\right)} \nonumber \\
 & &\times\frac{
\left\langle \mu _{2}\left|H_{E-L}^{+}(p)\right| \mu _{1}\right\rangle
\left\langle \mu _{1}\left|H_{E-R}^{-}\right| G\right\rangle }
{\left( \hbar \omega_{l}-E_{\mu _{1}}+i\Gamma _{\mu _{1}}\right)}
,  \label{eq:amplitude}
\end{eqnarray}
where $\Gamma _{\mu }$ are the lifetime broadenings of the electronic
excitations.\ The interaction matrix elements $\left\langle \mu _{2}\left|
H_{E-L}^{+}(p)\right| \mu _{1}\right\rangle $, $\left\langle G\left|
H_{E-R}^{+}\right| \mu _{2}\right\rangle $\ and $\left\langle \mu _{1}\left|
H_{E-R}^{-}\right| G\right\rangle $\ have been calculated under two
approximations of solid state theory: the Effective Mass Approximation and a
Long Wave Continuous Model for the optical phonons (a field theory
approach). Working expressions can be found elsewhere\cite%
{chamb,multifonon,fedorov,saqdprb,hraman,e1,e3}. The conditions of validity
of the above formalism can be summarized as: (1) The scattering process is
dominated by extended vibrational states, which are not affected by
single-electron excitations, as is usual in large molecules and solids; (2)
Excited electronic states are well separated in energy from the ground state.

\subsection{Albrecht's theory}

In this scheme, the wave functions of the molecule (or the quantum dot) are
considered in the adiabatic Born-Oppenheimer approximation 
\[
\Psi _{ev}(\{r\},\{Q\})=\Theta _{e}(\{r\},\{Q\})\Phi _{ev}(\{Q\}), 
\]%
where $e$ and $v$ are the sets of electronic and vibrational quantum
numbers, respectively; $\{r\}$ is the ensemble of electron coordinates of
the molecule (or the quantum dot), and $\{Q\}$ is the ensemble of normal
coordinates of the ions. The vibrational wave function $\Phi _{ev}$ is
factored as a product of the wave functions of all the normal modes 
\begin{equation}
\Phi _{ev}(\{Q\})=\varphi _{v_{1}}^{e}(Q_{1})\varphi
_{v_{2}}^{e}(Q_{2})\ldots ,  \label{ec:ev}
\end{equation}%
$v_{a}$ being phonon occupation numbers. According to the dispersion theory,
Albrecht obtained the Raman polarizability tensor for resonance scattering
as 
\[
\tensor{\bm{\alpha }}=\tensor{\bm{A}}^{\prime \prime \prime }+%
\tensor{\bm{B}}^{\prime \prime \prime }+\text{h.o.t.+n.r.t.}, 
\]%
where h.o.t. means higher order terms and n.r.t. non-resonant terms.

The A and B terms are given by 
\begin{equation}
\left( \tensor{\bm{A}}^{\prime \prime \prime }\right)
_{gi,gj}=\sum\limits_{(e),v}\bm{\mathfrak{M}}_{e,g}^{0}\bm{\mathfrak{M}}%
_{g,e}^{0}\frac{\left( gi|ev\right) \left( ev|gj\right) }{\hbar \omega
_{ev,gi}-\hbar \omega _{l}+i\Gamma _{e}^{\prime }}  \label{eq:A3term}
\end{equation}%
and 
\begin{widetext}
\begin{equation}
\left( \tensor{\bm{B}}^{\prime \prime \prime }\right)
_{gi,gj}=\sum\limits_{(e),v,s,a}\frac{h_{se}^{a}\bm{\mathfrak{M}}_{e,g}^{0}%
\bm{\mathfrak{M}}_{g,s}^{0}\left( gi|Q_{a}|ev\right) \left( ev|gj\right)
+h_{es}^{a}\bm{\mathfrak{M}}_{s,g}^{0}\bm{\mathfrak{M}}_{g,e}^{0}\left(
gi|ev\right) \left( ev|Q_{a}|gj\right) }{\left( \hbar \omega _{ev,gi}-\hbar
\omega _{l}+i\Gamma _{e}^{\prime }\right) \left( \hbar \omega _{e,s}\right) }%
.  \label{eq:B3term}
\end{equation}%
\end{widetext}
In the above expression we follow Albrecht's notation\cite{albrecht}, but we
use dyadic notation instead of tensor subscripts, $\hbar \omega _{l}$
instead of $h\nu _{0}$ for incoming photon energy, and $\hbar \omega
_{\alpha ,\beta }=E_{\alpha }-E_{\beta }$. $\left| gi\right) $, $\left|
ev\right) $, $\left| gj\right) $ are the initial, intermediate and final
vibrational states (Eq.~(\ref{ec:ev})), respectively, in the potential
energy fields of the electronic states $g$ (ground) and $e$ (excited). $%
\bm{\mathfrak{M}}_{\alpha ,\beta }^{0}$ and $h_{se}^{a}$ are, respectively,
dipole and electron-phonon interaction matrix elements (see Table\ \ref{tab1}%
). Also, the order of some matrix element indexes is reversed in order to
generalize the Albrecht's expressions for the case of complex wave functions.
Note that the order of subscripts in matrix elements in this notation is the
opposite of that in Dirac's notation. The summation on the index $(e)$ can
be restricted to the resonant state $e$. The $s$ states in (\ref{eq:B3term})
appear from a perturbative expansion of $\Theta _{e}(\{r\},\{Q\})$ in terms
of $\Theta _{s}(\{r\},\{0\})$.
\begin{table}
\caption{Equivalence of Albrecht and TDPT notations.\label{tab1}}
\begin{ruledtabular}
\begin{tabular}{p{2.3cm}cc}
Magnitude & Albrecht's & TDPT \\
\hline 
Electronic excited states & $e,s$ & $\mu _{2},\mu _{1}$ \\
Electronic ground state & $g$ & $G$ \\
Matrix element & $-i\sqrt{\frac{2\pi \hbar \omega _{s,g}}{V\eta _{l}^{2}}}%
{\bf e}_{l}\cdot \bm{\mathfrak{M}}_{g,s}^{0}$ & $\left\langle \mu _{1}\left|
H_{E-R}^{-}({\bf e}_{l},{\bf 0})\right| G\right\rangle $ \\
Matrix element & $x_{0}h_{se}^{p}$ & $\left\langle \mu _{2}\left|
H_{E-L}(p)\right| \mu _{1}\right\rangle $
\end{tabular}
\end{ruledtabular}
\end{table}                                                                    

The above formalism is valid if: (1) The $e$ states are non-degenerate or
uncoupled to other states with the same energy, and (2) Excited electronic
states are well separated in energy from the ground state.

\subsubsection{The offset oscillators model}

In this model, the vibrational states $\varphi _{i_{a}}^{g}(Q_{a})$ and $%
\varphi _{v_{a}}^{e}(Q_{a})$ are assumed to be localized in parabolic $%
V_{g(e)}(Q_{a})$ potentials with the same curvature, but the origin of $%
V_{e} $ shifted in $\Delta Q_{a}=\sqrt{2}x_{a0}\Delta _{a}=\sqrt{2}\sqrt{%
\hbar /\omega _{a}}\Delta _{a}$. In many cases, only one vibrational mode is
assumed and $\Delta _{a}$ is employed as fitting parameter. Moreover, $%
\Delta _{a}^{2}$ is identified as the Huang-Rhys factor\cite{klein,nomura}.
The integrals appearing in the A and B terms are given by 
\begin{equation}
\left( gl|ev\right) =\sqrt{\frac{v!}{l!}}e^{-\Delta _{a}^{2}/2}\Delta
_{a}^{l-v}L_{v}^{l-v}\left( \Delta _{a}^{2}\right) ,  \label{eq:osc_overlap}
\end{equation}
and 
\begin{eqnarray}
\left( ev|Q_{a}|gl\right) & = &x_{a0}\sqrt{\frac{l+1}{2}}\left(
g,l+1|ev\right) + \nonumber \\
 & & + x_{a0}\sqrt{\frac{l}{2}}\left( g,l-1|ev\right)
\nonumber \\
& = & x_{a0}\frac{l-v+\Delta _{a}^{2}}{\sqrt{2}\Delta _{a}}\left( gl|ev\right) ,
\label{eq:osc_Q}
\end{eqnarray}%
where $L_{m}^{p}$ are the Laguerre generalized polynomials and $x_{a0}=\sqrt{%
\hbar /\omega _{a}}$.

Several authors have used this model to study the electron-phonon coupling
in nanocrystals \cite{klein,jung,scamarcio,cucl,cubr,krauss2}. All of them
have considered only the term A and have used $\Delta _{a}^{2}$ to fit the
overtone/fundamental intensity ratios of Raman spectra. The fitted values of 
$\Delta _{a}^{2}$ are near 1, in contradiction with microscopic calculations%
\cite{nomura,chenjlum}.

\subsubsection{The limit of weak electron-phonon coupling}

In the limit of weak electron-phonon coupling the oscillator offset $\Delta
_{a}$ should be small. Expanding (\ref{eq:osc_overlap}) and (\ref{eq:osc_Q})
in powers of $\Delta _{a}$ we find 
\begin{equation}
\left( ev|gi\right) =\delta _{v,i}+\left[ \sqrt{\frac{i!}{v!}}\delta
_{v,i-1}-\sqrt{\frac{v!}{i!}}\delta _{v,i+1}\right] \Delta _{a}+O(\Delta
_{a}^{2})  \label{eq:evgi:weak}
\end{equation}%
and 
\begin{eqnarray}
\lefteqn{
\left( ev|Q_{a}|gi\right) =\int \varphi _{i_{a}}(Q_{a})\varphi
_{v_{a}}(Q_{a})Q_{a}\,dQ_{a} 
} \nonumber \\
& = &  x_{a0}\left[ \sqrt{\frac{v_{a}}{2}}\delta
_{v_{a},i_{a}+1}+\sqrt{\frac{i_{a}}{2}}\delta _{v_{a},i_{a}-1}\right]+ 
\frac{x_{0a}\Delta _{a}}{\sqrt{2}} \nonumber \\
& & \times\left[ \sqrt{\frac{i!}{v!}}\left(
(i+1)\delta _{v,i}+\delta _{v,i-2}\right) -\sqrt{\frac{v!}{i!}}\left(
i\delta _{v,i}+\delta _{v,i+2}\right) \right] \nonumber \\
& & +O(\Delta _{a}^{2}).
\label{eq:evQgi:weak}
\end{eqnarray}
Replacing (\ref{eq:evgi:weak})\ in Eq. (\ref{eq:A3term}) we obtain 
\[
\left( \tensor{
\bm{A}}^{\prime \prime \prime }\right) _{gi,g,j}=\delta _{i,j}\frac{%
\bm{\mathfrak{M}}_{e,g}^{0}\bm{\mathfrak{M}}_{g,e}^{0}}{E_{e,g}-\hbar \omega
_{l}-\Delta _{a}^{2}\hbar \omega _{a}+i\Gamma _{e}^{\prime }}+O(\Delta
_{a}), 
\]%
This means that $\tensor{
\bm{A}}^{\prime \prime \prime }$ contributes mainly to Rayleigh
scattering. Nevertheless, the term proportional to $\Delta _{a}$ is
important for one-phonon Raman scattering. To first order in $\Delta _{a}$
we find that 
\begin{widetext}
\[
\left( \tensor{
\bm{A}}^{\prime \prime \prime }\right) _{g0_{a},g1_{a}}=\frac{-%
\bm{\mathfrak{M}}_{e,g}^{0}\bm{\mathfrak{M}}_{g,e}^{0}\Delta _{a}\hbar \omega
_{a}}{\left( E_{e,g}-\Delta _{a}^{2}\hbar \omega _{a}-\hbar \omega
_{l}+i\Gamma _{e}^{\prime }\right) \left( E_{e,g}+\left( 1-\Delta
_{a}^{2}\right) \hbar \omega _{a}-\hbar \omega _{l}+i\Gamma _{e}^{\prime
}\right) }. 
\]

On the other hand, substituting (\ref{eq:evgi:weak})\ and (\ref%
{eq:evQgi:weak}) in (\ref{eq:B3term}) we obtain that, up to first order in
the electron-phonon interaction, the B term is non-null only for one-phonon
Raman processes. For Stokes processes at low temperature ($i_{a}=0$, $%
j_{a}=1 $) we obtain 
\begin{equation*}
\left( \tensor{\bm{B}}^{\prime \prime \prime }\right) _{g0_{a},g1_{a}}
=\sum\limits_{(e),s}\left\{ \frac{\bm{\mathfrak{M}}_{e,g}^{0}\bm{\mathfrak{M}}%
_{g,s}^{0}h_{se}^{a}x_{a0}/\sqrt{2}}{\left( E_{e,g}+\left( 1-\Delta
_{a}^{2}\right) \hbar \omega _{a}-\hbar \omega _{l}+i\Gamma _{e}^{\prime
}\right) \left( E_{e}-E_{s}\right) }\right.  
+\left. \frac{\bm{\mathfrak{M}}_{s,g}^{0}\bm{\mathfrak{M}}%
_{g,e}^{0}h_{es}^{a}x_{a0}/\sqrt{2}}{\left( E_{e,g}-\Delta _{a}^{2}\hbar
\omega _{a}-\hbar \omega _{l}+i\Gamma _{e}^{\prime }\right) \left(
E_{e}-E_{s}\right) }\right\} .
\end{equation*}
In the second term, one can exchange the indexes $e$ and $s$. Next, under
the condition $\left| E_{e}-E_{s}\right| \gg \left| \hbar \omega
_{a}+i(\Gamma _{e}^{\prime }-\Gamma _{s}^{\prime })\right| $ one obtains 
\begin{equation}
\left( \tensor{\bm{B}}^{\prime \prime \prime }\right)
_{g0_{a},g1_{a}}=\sum\limits_{e,s\neq e}\frac{-\bm{\mathfrak{M}}_{e,g}^{0}%
\bm{\mathfrak{M}}_{g,s}^{0}h_{se}^{a}x_{a0}/\sqrt{2}}{\left( E_{e,g}+\left(
1-\Delta _{a}^{2}\right) \hbar \omega _{a}-\hbar \omega _{l}+i\Gamma
_{e}^{\prime }\right) \left( E_{s,g}-\Delta _{a}^{2}\hbar \omega _{a}-\hbar
\omega _{l}+i\Gamma _{s}^{\prime }\right) }.  \label{eq:B3term:weak}
\end{equation}
\end{widetext} 
Noting that $\Delta _{a}^{2}\ll 1$, $\Delta _{a}\hbar \omega
_{a}=-\left\langle e\left| H_{E-L}(a)\right| e\right\rangle $, and $%
h_{se}^{a}x_{a0}/\sqrt{2}=\left\langle e\left| H_{E-L}(a)\right|
s\right\rangle $ (see Appendixes \ref{apendix2} and \ref{apendix3}), we see
that $\tensor{\bm{A}}^{\prime \prime \prime }+\tensor{\bm{B}}^{\prime
\prime \prime }$ for one-phonon emission reduces to the same result that the
TDPT.

\section{Calculation of $\tensor{\protect\bm{A}}^{\prime \prime \prime }$
and $\tensor{\protect\bm{B}}^{\prime \prime \prime }$}

\subsection{Effective Mass Approximation}

We use the exciton wave functions and the electron-phonon operator of Ref.~%
\onlinecite{e1} to estimate the A and B terms in nanocrystals of several
semiconductors. Using the Table \ref{tab1} and the Effective Mass
Approximation we have 
\[
\bm{\mathfrak{M}}_{g,e}^{0}=\frac{ie}{m_{0}\omega _{e,g}}{\bf p}%
_{s_{z}j_{z};e}f_{o_{e}} 
\]%
where $f_{o_{e}}$ is the envelope overlap integral ($o_{e}$ being the set of
envelope quantum numbers) 
\[
f_{o_{e}}=\int \Psi _{o_{e}}({\bf r},{\bf r})^{\ast }\ d^{3}{\bf r}. 
\]%
${\bf p}_{s_{z}J_{z};e}$ is the bulk momentum matrix element between the
couple of bands to which the exciton $e$ belongs. For the valence band $%
J_{z}=\pm 3/2,\pm 1/2$ and for the conduction band $s_{z}=\pm 1/2$.

\begin{table}
\caption{Parameters used in the calculations When not indicated, the source is
Ref.~\protect\onlinecite{e1} for CdS and Ref.~\protect\onlinecite{e2} for
CdSe.\label{tab3}}
\begin{ruledtabular}
\begin{tabular}{llll}
Parameter & CdS & CdSe & CdSe (MBEMA) \\ \hline
$E_{g}$ (eV) & 2.6 & 1.865 & 1.841\footnotemark[1] \\
$m_{e}/m_{0}$ & 0.18 & 0.12 & 0.13\footnotemark[2] \\
$m_{h}/m_{0}$ & 0.51 & 0.45 &  \\
$\gamma _{1}$ &  &  & 1.66\footnotemark[2] \\
$\gamma _{2}$ &  &  & 0.41\footnotemark[2] \\
$2m_{0}P^{2}$ (eV) & 21\footnotemark[3] & 20\footnotemark[3] & 20%
\footnotemark[3] \\
$\kappa $ & 7.8 & 9.53 & 9.53 \\
$V_{e}$ (eV) & 2.5 & $\infty $ & 0.6\footnotemark[2] \\ 
$V_{h}$ (eV) & 1.9 & $\infty $ & $\infty $\footnotemark[2] \\ 
$\omega _{L}$ (cm$^{-1}$) & 305 & 213 & 213 \\
$\omega _{T}$ (cm$^{-1}$) & 238 & 165 & 165 \\
$\epsilon _{0}$ & $8.7$\footnotemark[4] & 9.53 & 9.53 \\
$\epsilon _{\infty }$ & 5.3 & 5.72\footnotemark[4] & $5.72$\footnotemark[4]
\\
$\beta _{L}$ (10$^{-6}$) & 2.68 & 1.58 & 1.58 \\
$\Gamma _{\mu }$ (meV) & 5 & 5 & 5 \\
\end{tabular}
\end{ruledtabular}
\footnotetext[1]{Ref.~\onlinecite{landolt}.} \footnotetext[2]{Ref.~%
\onlinecite{laheld}.} \footnotetext[3]{Ref.~\onlinecite{hermann}.} %
\footnotetext[4]{Calculated from the Lydanne-Sachs-Teller relation.}
\end{table}                                                                   

We  evaluate the relative importance of the terms A and B for semiconductor
nanocrystals. Assuming only one vibrational mode and focusing on one-phonon
creation processes at low temperature (i. e. $i=0$ and $j=1$), the resonant
term in (\ref{eq:A3term}) is reduced to 
\begin{eqnarray}
\left( \tensor{\bm{A}}^{\prime \prime \prime }\right) _{gi,gj}=\left(
\sum_{s_{z},J_{z}}{\bf p}_{s_{z}J_{z};e}^{\ast }{\bf p}_{s_{z}J_{z};e}%
\right) \left( \frac{e\hbar f_{o_{e}}}{m_{0}E_{e,g}}\right)
^{2} \nonumber \\
\times\sum\limits_{v}\frac{\left( gi|ev\right) \left( ev|gj\right) }{%
E_{e,g}+\left( v-\Delta ^{2}\right) \hbar \omega _{LO}-\hbar \omega
_{l}+i\Gamma _{e}^{\prime }}.  \label{eq:A2calc}
\end{eqnarray}
The first term between parentheses is a band factor and is the responsible
of the angular pattern of the scattered intensity. The summation over $%
s_{z},J_{z}$, is performed to take into account the degeneracy of conduction
and hole bands, giving \cite{huangrhysn1} 
\begin{equation}
\sum_{s_{z},J_{z}}{\bf p}_{s_{z}J_{z};e}^{\ast }{\bf p}_{s_{z}J_{z};e}=\frac{%
(2J+1)}{3}\left( m_{0}P\right) ^{2}\tensor{\bm{1}},  \label{eq:band_sum}
\end{equation}%
where $J=3/2$ (1/2) if the upper valence band have $\Gamma _{8}$ ($\Gamma
_{7}$) symmetry and $P=-i\left\langle S\left| \hat{p}_{x}\right|
X\right\rangle /m_{0}$.

To evaluate the B term we also need to include the electron-phonon matrix
elements $x_{0}h_{se}=\sqrt{2}\left\langle e\left| H_{E-L}\right|
s\right\rangle $ (see Appendix \ref{apendix3}). With these considerations we
obtain 
\begin{eqnarray}
\label{eq:B2calc}
\lefteqn{
\left( \tensor{\bm{B}}^{\prime \prime \prime }\right) _{gi,gj}=\left(
\sum_{s_{z}J_{z}}{\bf p}_{s_{z}J_{z};e}^{\ast }{\bf p}_{s_{z}J_{z};s}\right)
\left( \frac{e\hbar }{m_{0}}\right) ^{2} 
} % close lefteqn environment  
\nonumber \\
& \times & \sum\limits_{v,os}\frac{f_{oe}f_{os}\left\langle e\left|
H_{E-L}\right| s\right\rangle }{E_{e,g}E_{s,g}E_{e,s}}
\frac{\sqrt{2}}{x_0} 
\nonumber \\
& & \times\frac{ \left( gi|Q|ev\right) \left(
ev|gj\right) +\left( gi|ev\right) \left( ev|Q|gj\right) }%
{E_{e,g}+\left( v-\Delta ^{2}\right) \hbar
\omega _{LO}-\hbar \omega _{l}+i\Gamma _{e}^{\prime } }
\end{eqnarray}

We evaluate $\left\langle e\left| H_{E-L}\right| s\right\rangle $ as in Ref. %
\onlinecite{e1}. Due to the Fr\"{o}hlich interaction cannot cause
intersubband transitions, ${\bf p}_{s_{z}J_{z};s}^{\ast }={\bf p}%
_{s_{z}J_{z};e}^{\ast }$ and the band factor in (\ref{eq:B2calc}) is the
same as that in (\ref{eq:A2calc}). Intersubband transitions may occur via
deformation potential interaction, but these are usually negligible in polar
materials. It must be noticed that it is not possible to consider this
mechanism within the Albrecht's theory, as the existence of degenerate hole
states connected by the electron-phonon interaction means a breakdown of the
adiabatic approximation, which is reflected in null denominators $E_{e,s}$
in Eq.~(\ref{eq:B2calc}). Nevertheless, TDPT can deal with it without
trouble. From this result we conclude that the A and B terms are scalars
(with small tensor corrections for B) and the integration over
nanocrystal orientations and light polarization has no effect on the ratio
between the A and B terms.

\subsection{Multiband effective mass theory}

In a multiband formalism, the essential effect of band mixing can be
captured using the spherical approximation for the hole Hamiltonian\cite%
{bald} 
\[
H_{h}=\frac{\gamma _{1}}{2m_{0}}\left( {\bf \hat{p}}^{2}-\frac{\mu }{9}%
\left( {\bf P}^{(2)}\cdot {\bf J}^{(2)}\right) \right) +V(r), 
\]%
$V(r)$ being the confinement potential, ${\bf P}^{(2)}$ and ${\bf J}^{(2)}$
are spherical rank tensors built from linear and angular momentum operators, 
$\mu =2\gamma _{2}/\gamma _{1}$, and $\gamma _{2}$ and $\gamma _{1}$ are
Luttinger parameters.

Electron-hole pair states with well defined total (Bloch+orbital) angular
momentum quantum numbers $M$ and $M_{z}$ can be obtained as 
\begin{eqnarray*}
\left| e\right\rangle &=&\sum_{n,N,l,L,f,F}  C_{nNlLsJfFMM_{z}}^{(e)}\left|
nNlLsJfFMM_{z}\right\rangle \\
&=&\sum_{n,N,l,L,f,F,f_{z},F_{z}} C_{nNlLsJfFMM_{z}}^{(e)}\left(
fFf_{z}F_{z}|MM_{z}\right) \\
 & & \qquad\qquad\qquad\times\left| nlsff_{z}\right\rangle \otimes \left|
NLJFF_{z}\right\rangle  ,
\end{eqnarray*}%
where $\left( fFf_{z}F_{z}|MM_{z}\right) $ is a Clebsch-Gordan coefficient.
Lowercase (uppercase) letters denote electron (hole) quantum numbers. $%
\left| nlsff_{z}\right\rangle $ and $\left| NLJFF_{z}\right\rangle $ are
electron and hole states with well defined total angular momentum, their
wave functions given by 
\begin{equation*}
\left\langle {\bf r}|nlsff_{z}\right\rangle =\sum_{l_{z},s_{z}}\left(
lsl_{z}s_{z}|ff_{z}\right) R_{nl}(r)Y_{ll_{z}}(\theta ,\varphi )\left\langle 
{\bf r}|ss_{z}\right\rangle ,
\end{equation*}
and
\begin{eqnarray*}
\left\langle {\bf r}|NLJFF_{z}\right\rangle 
&=&\sum_{K=L,L+2}\sum_{L_{z},J_{z}}\left( KJL_{z}J_{z}|ff_{z}\right) \\
& &  \times
R_{NK}^{(F,L)}(r)Y_{KL_{z}}(\theta ,\varphi )\left\langle {\bf r}%
|JJ_{z}\right\rangle .
\end{eqnarray*}%
In the above expression $R_{nl}(r)$ are the radial wave functions of a
particle in a spherical box and $R_{NK}^{(F,L)}(r)$ are the solutions of the
MBEMA equations given elsewhere\cite{baixia,efros92}, $Y_{ll_{z}}(\theta
,\varphi )$ are the spherical harmonics\cite{jackson}, $\left\langle {\bf r}%
|ss_{z}\right\rangle $ are $\Gamma _{6}$ Bloch function and $\left\langle 
{\bf r}|JJ_{z}\right\rangle $ are hole Bloch functions. The hole Bloch
functions are related with the $\Gamma _{8}$ ($J=3/2$) electronic Bloch
functions $\left| \overline{J,J_{z}}\right\rangle $ by the rule $\left|
JJ_{z}\right\rangle =(-1)^{J-J_{z}}\left| \overline{J,-J_{z}}\right\rangle $%
\ (derived from the time-reversal operation). Our $\left|
JJ_{z}\right\rangle $ are $\left| 3/2,\pm 3/2\right\rangle =\mp (i/\sqrt{2}%
)(X\pm iY)\left| \pm \right\rangle $, and $\left| 3/2,\pm 1/2\right\rangle
=(i/\sqrt{6})\left[ 2Z\left| \pm \right\rangle \mp (X\pm iY)\left| \mp
\right\rangle \right] $.

Within this basis, in the strong confinement regime, the Coulomb interaction
can be treated by direct diagonalization of the Hamiltonian or even by
simple perturbation theory. Using the theory of angular momentum \cite{brink}%
, compact expressions for the matrix elements can be obtained. The dipole
matrix elements gives 
\begin{eqnarray*}
\lefteqn{
\bm{\mathfrak{M}}_{g,e}^{0} =i{\bf \hat{e}}_{M_{z}}^{\ast }\frac{2Pe\hbar }{%
E_{e,g}}\delta _{M,1}(-1)^{f+5/2}\sqrt{\frac{(2f+1)(2F+1)}{3}} } \\
& & \times\left\{ 
\begin{array}{ccc}
1 & 3/2 & 1/2 \\ 
l & f & F%
\end{array}%
\right\} \left( \delta _{l,L}+\delta _{l,L+2}\right) \int
R_{Nl}^{(F,L)}(r)R_{nl}(r)r^{2}dr,
\end{eqnarray*}%
where $R_{nl}$ and $R_{NL}^{(F,L)}$ are electron and hole radial functions,
respectively, and 
\[
{\bf \hat{e}}_{0}={\bf k},\qquad {\bf \hat{e}}_{\pm 1}=\mp \frac{{\bf i}\pm i%
{\bf j}}{\sqrt{2}}, 
\]
$\bf{i}$, $\bf{j}$ and $\bf{k}$ being the unit vectors along the X-, Y- 
and Z-axis, respectively.

In the A term we must sum over the degenerate $e$ states with different $%
M_{z}$, which turns out in a term proportional to the diagonal tensor 
\begin{widetext}
\begin{eqnarray*}
\left( \tensor{\bm{A}}^{\prime \prime \prime }\right) _{gi,gj} &=&%
\tensor{\bm{1}}\frac{4P^{2}e^{2}\hbar ^{2}}{E_{e,g}^{2}}\delta
_{M,1}\left( \delta _{l_{e},L_{e}}+\delta _{l_{e},L_{e}+2}\right) \frac{%
(2f_{e}+1)(2F_{e}+1)}{3}\left\{ 
\begin{array}{ccc}
1 & 3/2 & 1/2 \\ 
l_{e} & f_{e} & F_{e}%
\end{array}%
\right\} ^{2} \\
&&\times \left[ \int
R_{N_{e}l_{e}}^{(F_{e},L_{e})}(r)R_{n_{e}l_{e}}(r)r^{2}dr\right] ^{2}\sum_{v}%
\frac{(gi|ev)(ev|gi)}{E_{e,g}+\left( v-\Delta ^{2}\right) \hbar \omega
_{LO}-\hbar \omega _{l}+i\Gamma _{e}^{\prime }}.
\end{eqnarray*}

The B term for $l_{p}=0$ phonons is given by 
\begin{eqnarray*}
\left. \tensor{\bm{B}}^{\prime \prime \prime }\right| _{l_{p}=0} &=&%
\tensor{\bm{1}}\delta _{M,1}\sum_{(e),s}\frac{4P^{2}e^{2}\hbar ^{2}}{%
3E_{e,g}E_{s,g}}\left\{ 
\begin{array}{ccc}
1 & 3/2 & 1/2 \\ 
l_{e} & f_{e} & F_{e}%
\end{array}%
\right\} \left\{ 
\begin{array}{ccc}
1 & 3/2 & 1/2 \\ 
l_{s} & f_{s} & F_{s}%
\end{array}%
\right\} (-1)^{f_{e}+f_{s}+3} \\
&&\times \sqrt{(2f_{e}+1)(2f_{s}+1)(2F_{e}+1)(2F_{s}+1)}
\left( \delta _{l_{e},L_{e}}+\delta
_{l_{e},L_{e}+2}\right) \left( \delta _{l_{s},L_{s}}+\delta
_{l_{s},L_{s}+2}\right) \\
&&\times \int R_{N_{e}l_{e}}^{(F,L_{e})}(r)R_{n_{e}l_{e}}(r)r^{2}dr\int
R_{N_{s}l_{s}}^{(F,L_{s})}(r)R_{n_{s}l_{s}}(r)r^{2}dr\ \left\langle s\left|
H_{E-L}^{(l_{p}=0)}\right| e\right\rangle \\
&&\times \sum_{v}\frac{\sqrt{2}}{x_{0}}\frac{\left( gi|Q|ev\right) \left(
ev|gj\right) +\left( gi|ev\right) \left( ev|Q|gj\right) }{\left(
E_{e,s}\right) \left( E_{e,g}+(v-i-\Delta ^{2})\hbar \omega _{LO}-\hbar
\omega _{l}+i\Gamma _{e}^{\prime }\right) .}
\end{eqnarray*}
\end{widetext}
\section{Discussion}

Table~\ref{tab2} shows the Huang-Rhys parameter $\Delta ^{2}$, the A term
absolute value, and the ratio $\left| A^{\prime \prime \prime }/B^{\prime
\prime \prime }\right| $, calculated for typical nanocrystals 20~\AA\ in
radius. The photon energies, corresponding to incoming resonance with the
lower exciton level are also indicated in the table. The incomplete exciton
confinement in CdS nanocrystals has been considered with two different
models: (1) Finite band offsets $V_{e}$ and $V_{h}$\cite{e3}, and (2) An
effective radius\cite{e1}. Moreover, for CdSe nanocrystals we have also
considered two models: (1) Effective radius\cite{e2} and (2) Multiband
effective mass approximation (MBEMA) along with finite conduction band
offset $V_{e}$\cite{laheld}.

\begin{table}
\caption{Numerical results for typical QD's 20 \AA{} in radius. The energies $\hbar\omega_l$
correspond to incoming resonance with the lower Raman active exciton level
in each nanocrystal. \label{tab2}}
\begin{ruledtabular}
\begin{tabular}{lllll}
Nanocrystal & $\Delta ^{2}$ & $\hbar \omega _{l}$\ (eV) & $\left| A^{\prime
\prime \prime }/B^{\prime \prime \prime }\right| $ & $\left| A^{\prime
\prime \prime }\right| $\ (\AA $^{3}$) \\ \hline
CdS & $0.08$ & $2.870$ & $13$ & $8.1\times 10^{4}$ \\
CdS ($R_{ef}$) & $0.0013$ & $2.878$ & $8.4$ & $1.2\times 10^{4}$ \\
CdSe & $0.0008$ & $2.592$ & $27$ & $7.9\times 10^{3}$ \\
CdSe (MBEMA) & $0.2$ & $2.280$ & $32$ & $6.4\times 10^{4}$%
\end{tabular}
\end{ruledtabular}
\end{table}   

As it has been noticed\cite{e1,e3}, Raman scattering is quasi-forbidden in
quantum dots in the strong confinement regime. The scattering is possible
through a {\em decompensation} between the electron and hole wave functions,
which may appear due to (1) the Coulomb electron-hole interaction and
difference between electron and hole masses, (2) difference in electron and
hole confinement, (3) hole band mixing, (4) defects, and (5) non-adiabatic
effects\cite{pokatilov01}. The electron-hole decompensation also has a
direct relation with the Huang-Rhys parameter. This one is larger in the
case of CdS, where incomplete confinement have been considered, and in CdSe
when band mixing and incomplete electron confinement have been included.
These ones are the cases where larger Raman polarizabilities are obtained.
Notice that, due to the finite electron confinement assumed, our Huang-Rhys
parameter for CdSe nanocrystals within MBEMA is larger than other reported
theoretical values\cite{nomura,chenjlum} and is within the order of
magnitude of the experimental values\cite%
{klein,alivisatos,bawendi90,mittleman}.

In all the cases examined, the A term determines the Raman polarizability.
This means that the interpretation of one-phonon Raman cross section
considering only the A term is consistent with the microscopic calculations
using TDPT reported here and in Refs.~\onlinecite{chamb,e1,e3}. However, it
is incorrect to fit the Huang-Rhys factor from the overtone to fundamental
intensity ratios in multiphonon Raman spectra, as different scattering
channels give substantial contributions to the overtones \cite%
{pokatilov01,multifonon,fedorov}.

Let us consider the participation of optical phonons with $l_{p}>0$ in Raman
scattering. These phonons connect degenerate band-mixed exciton states and
breaks the adiabatic approximation. Hence, the Albrecht's theory cannot
describe Raman scattering from these phonons. On the other hand, the TDPT,
may deal with degenerate states and non-adiabatic processes. Calculations in
Ref.~\onlinecite{efros91} indicate that the role of $l_{p}>0$ phonons in
one-phonon Raman spectra is to cause a small shoulder near the interface
phonon frequency, being unimportant for the principal peak. Nevertheless,
the electron-lattice interaction breaks the degeneracy of the exciton
states, causing a redistribution of the exciton-phonon energy levels and
possibly originate exciton-phonon complexes. Both these factors could
substantially alter the predicted Raman spectra. This effect can be
considered by TDPT of higher order in the electron-lattice interaction, or
by a non-perturbative calculation of the exciton-phonon complexes. Research
on this direction is presently in progress.

\section{Conclusions}

\label{conclusion}

We have established the connection between the two theories more used for
resonance Raman scattering in semiconductor nanocrystals: the Albrecht's
theory based on the adiabatic approximation and the time dependent
perturbative approach. In particular, we have shown that both theories are
equivalent in the limit of weak electron-lattice interaction and when the
resonant exciton level is non-degenerate or is a set of uncoupled degenerate
states. We have evaluated the relative importance of the Albrecht's A and B
terms (the last one not discussed in the literature) for CdSe and CdS
nanocrystals, using different models for the electronic excitations. We have
found that the A term is the leading coefficient in all the cases
considered. Additionally, we have given the expressions of the matrix
elements of the electron-radiation and the electron-lattice interactions for
a model of exciton considering the fourfold degeneracy of the $\Gamma _{8}$
valence band. In the framework of this model, we have obtained a theoretical
Huang-Rhys parameter within the order of magnitude of experimental values.

\acknowledgments

The authors are grateful to Prof. Carlos Trallero-Giner for their useful
suggestions and for a critical reading of the manuscript. This work was
partially supported by Alma Mater project 26-2000 of Havana University.

\appendix

\section{The oscillator offset}

\label{apendix2}

The adiabatic vibrational eigenstates for the electronic state $e$ are
obtained from the Hamiltonian 
\begin{eqnarray*}
\hat{H}_{vibr}^{e} &=&\left\langle e\left| H_{L}+H_{E-L}\right|
e\right\rangle \\
&=&\sum_{a}\left\{ \hbar \omega _{a}\hat{b}_{a}^{\dagger }\hat{b}%
_{a}+\left\langle e\left| H_{E-L}(a)\right| e\right\rangle \left( \hat{b}%
_{a}+\hat{b}_{a}^{\dagger }\right) \right\} .
\end{eqnarray*}%
We have considered the case of Hermitian $H_{E-L}(a)$. For non-Hermitian $%
H_{E-L}(a)$ see Appendix~\ref{apendix3}. The unitary transformation $\hat{c}%
_{a}=\hat{b}_{a}-\alpha _{a}$, with $\alpha _{a}=-\left\langle e\left|
H_{E-L}(a)\right| e\right\rangle /\hbar \omega _{a}$ diagonalizes the
Hamiltonian 
\[
H_{vibr}^{e}=\sum_{a}\hbar \omega _{a}(\hat{c}_{a}^{\dagger }\hat{c}%
_{a}-\alpha _{a}^{2}). 
\]%
The new and old phonon coordinate operators are related by 
\[
Q_{a}^{e}=\sqrt{\frac{\hbar }{2\omega _{a}}}\left( \hat{c}_{a}+\hat{c}%
_{a}^{\dagger }\right) =Q_{a}^{G}-\sqrt{2}\sqrt{\frac{\hbar }{\omega _{a}}}%
\alpha _{a}. 
\]%
This relation identifies $\alpha _{a}$ with the oscillator offset $\Delta
_{a}$.

\section{The connection between the classical and the quantum
electron-phonon interaction}

\label{apendix3} The Hamiltonian operator determining the electronic
eigenstates can be expanded in Taylor series of the vibrational normal
coordinates 
\[
H_{E}(Q)=H_{E}(0)+\sum_{a}\frac{\partial H_{E}}{\partial Q_{a}}Q_{a}.
\]%
The second term of the above formula is the electron-phonon interaction. 
\begin{equation}
H_{E-L}=\sum_{a}\frac{\partial H_{E}}{\partial Q_{a}}Q_{a}.
\label{eq:hel:classic}
\end{equation}%
The quantum Fr\"{o}hlich-type interaction operator have the form\cite{phil} 
\[
H_{E-L}=\sum_{a}H_{E-L}^{-}(a)\hat{b}_{a}+H_{E-L}^{+}(a)\hat{b}_{a}^{\dagger
},
\]%
where $\hat{b}_{a}$ are annihilation operator of phonons in the normal modes 
$a$. $H_{E-L}^{\pm }(a)$ are operators that act on the electronic
coordinates, e.g. $H_{E-L}^{-}(n,l,m)=\Phi _{n,l}(r)Y_{l,m}(\theta ,\varphi )
$ for the one-electron-phonon interaction in a semiconductor nanocrystal %
\cite{chamb}. Making the transformation to coordinate and momentum operators 
\[
\hat{b}_{a}=\sqrt{\frac{\omega _{a}}{2\hbar }}\left( Q_{a}+\frac{i}{\omega
_{a}}P_{a}\right) ,
\]%
the interaction operator becomes 
\begin{eqnarray}
H_{E-L} &=&\sum_{a}\sqrt{\frac{\omega _{a}}{2\hbar }}\left(
H_{E-L}^{-}(a)+H_{E-L}^{+}(a)\right) Q_{a} \nonumber \\
& & {}+\frac{i}{\sqrt{2\hbar \omega _{a}}%
}\left( H_{E-L}^{-}(a)-H_{E-L}^{+}(a)\right) P_{a}. \quad 
\label{eq:hel:quantum}
\end{eqnarray}%
If $H_{E-L}^{-}(a)=H_{E-L}^{+}(a)=H_{E-L}(a)$ then (\ref{eq:hel:classic})
and (\ref{eq:hel:quantum}) are equivalent and 
\[
h_{es}^{a}=\left\langle s\left| \frac{\partial H_{el}}{\partial Q_{a}}%
\right| e\right\rangle =\sqrt{\frac{2\omega _{a}}{\hbar }}\left\langle
s\left| H_{E-L}(a)\right| e\right\rangle .
\]%
When the vibrational modes in a nanostructure are described by complex
fields there are modes for which $H_{E-L}^{-}(a)\neq H_{E-L}^{+}(a)$. In
this case, thanks to time reversal symmetry, complex modes are double
degenerate and real fields can be obtained from the real and imaginary parts
of the complex fields, which correspond to real normal coordinates. The new
matrix elements can be obtained from the complex matrix elements as 

{\setlength\arraycolsep{1pt}
\begin{subequations}
\label{eq:normal:real}
\begin{eqnarray}
\left\langle s\left| H_{E-L}^{(1)}(a)\right| e\right\rangle  &=&\frac{%
\left\langle s\left| H_{E-L}^{-}(a)+H_{E-L}^{+}(a)\right| e\right\rangle }{%
\sqrt{2}}, {}\\
\left\langle s\left| H_{E-L}^{(2)}(a)\right| e\right\rangle  &=&i\frac{%
\left\langle s\left| H_{E-L}^{-}(a)-H_{E-L}^{+}(a)\right| e\right\rangle }{%
\sqrt{2}} . {} 
\end{eqnarray}
\end{subequations}
}
\begin{widetext}

\section{Exciton phonon matrix elements with degenerate bands}

The matrix elements of the interaction of band-mixed excitons with the
optical phonons in a spherical nanocrystal can be calculated following the
procedure outlined in Ref.~\onlinecite{hraman}. We obtained the expression 
%\begin{widetext}
\begin{eqnarray*}
\lefteqn{
 \left\langle n^{\prime }N^{\prime }l^{\prime }L^{\prime }sJf^{\prime
}F^{\prime }M^{\prime }M_{z}^{\prime }\left|
H_{E-L}^{-}(n_{p},l_{p},m_{p})\right| nNlLsJfFMM_{z}\right\rangle
        } \\
& = &\left\langle nNlLsJfFMM_{z}\left| H_{E-L}^{+}(n_{p},l_{p},m_{p})\right|
n^{\prime }N^{\prime }l^{\prime }L^{\prime }sJf^{\prime }F^{\prime
}M^{\prime }M_{z}^{\prime }\right\rangle \\
& = &(-1)^{M^{\prime }-M_{z}^{\prime }}\frac{C_{F}}{\sqrt{R}}\left( 
\begin{array}{ccc}
M^{\prime } & l_{p} & M \\ 
-M_{z}^{\prime } & m_{p} & M_{z}%
\end{array}%
\right) \sqrt{(2M+1)(2M^{\prime }+1)(2l_{p}+1)/4\pi } \\
 & & \times \left\{ -\delta _{F,F^{\prime }}\delta _{N,N^{\prime }}\delta
_{L,L^{\prime }}(-1)^{F+M+f+f^{\prime }+s}\sqrt{(2f^{\prime
}+1)(2f+1)(2l^{\prime }+1)(2l+1)}\right. \\
 & &\quad \left. \times \left\{ 
\begin{array}{ccc}
f^{\prime } & f & l_{p} \\ 
l & l^{\prime } & s%
\end{array}%
\right\} \left\{ 
\begin{array}{ccc}
M^{\prime } & M & l_{p} \\ 
f & f^{\prime } & F%
\end{array}%
\right\} \left( 
\begin{array}{ccc}
l^{\prime } & l_{p} & l \\ 
0 & 0 & 0%
\end{array}%
\right) \int R_{n^{\prime }l^{\prime }}(r)\Phi
_{n_{p},l_{p}}(r)R_{nl}(r)r^{2}dr\right. \\
& & \quad \left. +\delta _{f,f^{\prime }}\delta _{n,n^{\prime }}\delta
_{l,l^{\prime }}(-1)^{M^{\prime }+f+J+2F}\sum_{K=L,L+2}\sum_{K^{\prime
}=L^{\prime },L^{\prime }+2}\sqrt{(2F^{\prime }+1)(2F+1)(2K^{\prime
}+1)(2K+1)}\right. \\
& & \quad \left. \times \left\{ 
\begin{array}{ccc}
F^{\prime } & F & l_{p} \\ 
K & K^{\prime } & J%
\end{array}%
\right\} \left\{ 
\begin{array}{ccc}
M^{\prime } & M & l_{p} \\ 
F & F^{\prime } & f%
\end{array}%
\right\} \left( 
\begin{array}{ccc}
K^{\prime } & l_{p} & K \\ 
0 & 0 & 0%
\end{array}%
\right) \int R_{N^{\prime }K^{\prime }}^{(F^{\prime },L^{\prime })}(r)\Phi
_{n_{p},l_{p}}(r)R_{NK}^{(F,L)}(r)r^{2}dr\right\} ,
\end{eqnarray*}
where $\Phi _{n_{p},l_{p}}(r)$ is the radial part of the optical phonons
electrostatic potential \cite{chamb}. The optical modes for $l_{p}>0$ are
described by complex fields, real field matrix elements can be obtained from
Eq.~(\ref{eq:normal:real}).

For the Coulomb electron-hole interaction we obtained the expression 
\begin{multline*}
\left\langle n^{\prime }N^{\prime }l^{\prime }L^{\prime }sJf^{\prime
}F^{\prime }M^{\prime }M_{z}^{\prime }\left| \frac{1}{\left| {\bf r}_{e}-%
{\bf r}_{h}\right| }\right| nNlLsJfFMM_{z}\right\rangle =\delta
_{M,M^{\prime }}\delta _{M_{z},M_{z}^{\prime }}(-1)^{s+J+2f+F+F^{\prime }+M}
\\
\times \sqrt{(2f+1)(2f^{\prime }+1)(2F+1)(2F^{\prime }+1)}\sum_{K,K^{\prime
},p}\left\{ 
\begin{array}{ccc}
p & f & f^{\prime } \\ 
s & l^{\prime } & l%
\end{array}%
\right\} \left\{ 
\begin{array}{ccc}
p & F & F^{\prime } \\ 
J & K^{\prime } & K%
\end{array}%
\right\} \left\{ 
\begin{array}{ccc}
p & f & f^{\prime } \\ 
M & F^{\prime } & F%
\end{array}%
\right\} \\
\times \sqrt{(2l+1)(2l^{\prime }+1)(2K+1)(2K^{\prime }+1)}\left( 
\begin{array}{ccc}
l^{\prime } & p & l \\ 
0 & 0 & 0%
\end{array}%
\right) \left( 
\begin{array}{ccc}
K^{\prime } & p & K \\ 
0 & 0 & 0%
\end{array}%
\right) \\
\times \int \int R_{n^{\prime }l^{\prime }}(r_{e})R_{nl}(r_{e})R_{N^{\prime
}K^{\prime }}^{(F^{\prime },L^{\prime })}(r_{h})R_{NK}^{(F,L)}(r_{h})\frac{%
r_{<}^{p+2}}{r_{>}^{p-1}}dr_{e}dr_{h}.
\end{multline*}%
In the above expression, due to the properties of the 3j-symbols, the
summation on $p$ runs from $\max (\left| l-l^{\prime }\right| ,\left|
K-K^{\prime }\right| )$ to $\min (l+l^{\prime },K+K^{\prime })$.
\end{widetext}

\bibliography{refer} % uses refer.bib

\end{document}